# A Hybrid Intelligent Framework for Maximising SAG Mill Throughput: An Integration of Expert Knowledge, Machine Learning and Evolutionary Algorithms for Parameter Optimisation

Zahra Ghasemi[a,*], Mehdi Neshat[b,h], Chris Aldrich[c], John Karageorgos[d], Max Zanin[e,f], Frank Neumann[g], Lei Chen[a]

[a] *School of Electrical and Mechanical Engineering, The University of Adelaide, North Terrace, Adelaide, SA 5005, Australia*

[b] *Centre for Artificial Intelligence Research & Optimisation, Torrens University Australia, Brisbane 4006, Australia,*

[c] *Western Australian School of Mines, Curtin University, Perth, Western Australia 6845, Astralia*

[d] *Manta Controls Pty Ltd, 1 Sharon Pl, Grange, SA 5022, Australia*

[e] *School of Chemical Engineering, The University of Adelaide, North Terrace, Adelaide, SA 5005, Australia*

[f] *School of Civil, Environmental and Mining Engineering, The University of South Australia, Adelaide, SA, 5095,Australia.*

[g] *School of Computer and Mathematical Sciences, The University of Adelaide, North Terrace, Adelaide, SA 5005, Australia.*

[h] *Faculty of Engineering and Information Technology, The University of Technology Sydney, Ultimo, NSW 2007, Australia*

**Abstract**

In mineral processing plants, grinding is a crucial step, accounting for approximately 50% of the total mineral processing costs. Semi-autogenous grinding (SAG) mills are extensively employed in the grinding circuit of mineral processing plants. Maximising SAG mill throughput is of significant importance considering its profound financial outcomes. However, the optimum process parameter setting aimed at achieving maximum mill throughput remains an uninvestigated domain in prior research. This study introduces a hybrid intelligent framework leveraging expert knowledge, machine learning techniques and evolutionary algorithms to address this research need. In this study, we utilise an extensive industrial dataset comprising 36,743 records and select relevant features based on the insights of industry experts. Following the removal of erroneous data, a comprehensive evaluation of 17 diverse machine learning models is undertaken to identify the most accurate predictive model. To improve the model's performance, feature selection and outlier detection are executed. The resultant optimal model, trained with refined features, serves as the objective function within three distinct evolutionary algorithms. These algorithms are employed to identify parameter configurations that maximise SAG mill throughput while adhering to the working limits of input parameters as constraints. Notably, our analysis revealed that CatBoost, as an ensemble model, stands out as the most accurate predictor. Furthermore, differential evolution emerges as the preferred optimisation algorithm, exhibiting superior performance in both achieving the highest mill throughput predictions and ensuring robustness in predictions, surpassing alternative methods.

## 1. Introduction

The process of separating valuable minerals from waste materials in mineral processing plants involves a series of intricate procedures. These procedures can be categorised generally into crushing, grinding, and concentration. Grinding is one of the most important procedures, where unconnected media such as balls, rods, or pebbles are utilised for particle size reduction. This process is usually performed wet to produce slurry for the concentration step. This operation is the most energy-intensive stage, constituting around 50 percent of all mineral processing costs [1].

Mineral processing plants commonly employ semi-autogenous grinding (SAG) mills for size reduction within their grinding circuits. Despite autogenous grinding (AG) mills, which rely exclusively on the ore for grinding purposes, SAG mills utilise a combination of steel balls and coarse ore to achieve the desired discharge particle size [2]. An essential performance metric for a SAG mill is its throughput, which measures the volume of processed ore per hour [3]. SAG mill throughput maximisation stands as a critical priority, given its profound financial outcomes. However, achieving this goal is a complex task, as numerous factors impact mill throughput. Moreover, the relationships between these factors and mill throughput are highly nonlinear, and input factors can mutually influence each other.

In some research studies, the actual operations of a mine site were investigated to explore how the mill throughput can be increased. For this purpose, some practical experiments or simulations were performed. Behnamfard et al. [4] reported that by adding large lumps of ore in mill feed, mill throughput can be increased. According to their experiments, a 30% improvement in the mill throughput resulted from raising the percentage of coarse material (larger than 200 mm) from 10% to 30%. In contrast, an increase from 30% to 45% had no beneficial effect on the mill throughput, and percentages higher than 45% even had a detrimental impact. In another study, the relationship between turning speed, mill filling, and mill throughput was investigated [5]. Their experiments revealed that depending on the turning speed, mill throughput peaks in fillings between 23% and 34%. Moreover, faster turning speeds resulted in higher mill throughputs. Another research in the Antamina mining complex in Peru using JKSimMet simulator revealed that increasing fine particles, which leads to lower F80, will increase mill throughput [6]. In another research for the Cadia Hill SAG mill circuit, throughput increased by increasing the SAG mill ball charge, reducing the number of lifter rows, reducing the rock charge, and reducing the SAG mill feed size distribution from an average F80 of 100mm to 80mm. These research efforts present interesting findings for increasing mill throughput while also demonstrating the complex relationship between inputs and mill throughput. However, these studies often necessitate real-world experiments, which require substantial financial and time dedication. Furthermore, they don't offer a general framework for maximising mill throughput through process parameter optimisation.

Recent advancements in sensor technology combined with improvements in data recording and storage methods have substantially increased the volume of production data stored in mineral processing plant databases. Advanced machine learning (ML) methods and artificial intelligence (AI) techniques can be employed to leverage these rich data resources for the purpose of maximising SAG mill throughput. For this purpose, an accurate prediction model needs to be developed using machine learning techniques. Subsequently, AI-based optimization algorithms will be employed to identify the optimal feature settings that can lead to maximum mill throughput. This comprehensive approach leverages both the wealth of available data and the computational power of ML and AI techniques to assist the mining industry in achieving enhanced and intelligent process control with the objective of maximising mill throughput.

The application of ML approaches for mill throughput prediction has been the subject of a few prior research studies. Both Dimitrakopoulos [7] applied neural network (NN) and multiple linear regression (MLR) models to predict ball mill throughput at the Tropicana Gold mining complex. They considered the average penetration rate from blast hole drilling as the ore hardness indicator, along with ball mill power, feed, and product particle sizes, as effective features for predicting ball mill throughput. They found that incorporating feed particle size data does not considerably increase model accuracy. Consequently, they did not include it in the final modelling. Ultimately, NN outperformed MLR. In this research, they could accurately model ball mill throughput, with RMSE 28.15 tonnes per hour. However, the study relied on a limited dataset comprising only 181 records. Moreover, despite the wide array of machine learning techniques available for predictive modelling, another limitation of the study is the exclusive comparison of just two approaches. Ghasemi et al. [8] compared six ML models, including genetic programming, recurrent neural networks, support vector regression, regression trees, random forest regression, and linear regression, for predicting SAG mill throughput. They utilised an industrial data set comprising 20,161 records, with the SAG mill's power draw, feed particle size, inlet water, and turning speed identified as effective features. To enhance prediction performance, delays in the data were identified, and hyperparameter tuning was conducted to determine the optimal settings for each model before implementation. The comparative analysis revealed that the recurrent neural network demonstrates the highest accuracy in predicting SAG mill throughput, followed by genetic programming and support vector regression. However, in their predictive framework, ensemble models were not considered.

While the aforementioned studies focused on modelling mill throughput using ML techniques, identifying the optimal set of control parameters for achieving the maximum mill throughput remains relatively unexplored in the current scientific literature. The current research aims to address this gap by providing an integrated and intelligent framework comprised of prediction and optimisation models using ML and AI techniques. This study utilises an industrial data set of 36,743 records from a gold mining complex in Western Australia, and twenty features are selected at the initial stage based on industry expert insight, including mill weight, power draw, turning speed, inlet water, feeder ratios, pebble crusher working status, and feed particle size. A comprehensive set of 17 ML models are compared to identify the most accurate prediction model. These models are selected from a wide range of machine learning prediction models, which include ensemble models, neural network models, tree-based models, and traditional machine learning models. The most accurate prediction model is then identified, and feature selection and outlier detection techniques are implemented to increase the prediction accuracy. Following these steps, the performance of three distinct evolutionary algorithms for optimisation, including PSO, DE, and GA are compared employing the best prediction model.

The subsequent sections of this study are organised as follows. The next section provides some details about the mining process along with descriptive statistics of the features. Section 3 describes the methods used in this study. Results are provided in section 4, and finally, the research's findings are summarised in section 5.

## 2. Process description and data analysis

The dataset employed in this research comprises 36,743 operational records from a gold mining complex. Table 1 shows the descriptive statistics of the cleaned data set. Figure 2 illustrates the histogram plot of all features, respectively.

Table 1: Basic descriptive statistics of the data

| Variable | Unit | Min | Max | Mean | Std. Deviation |
|---|---|---|---|---|---|
| Mill throughput | t/h | 0.00 | 1616.99 | 1202.70 | 136.18 |
| Mill weight | t | 440.14 | 744.59 | 654.66 | 55.03 |
| Mill power draw | kW | 0.01 | 14424.22 | 13469.80 | 711.19 |
| Mill turning speed | rpm | 0.00 | 10.31 | 9.91 | 0.47 |
| Inlet water | m³/h | 7.40 | 482.86 | 364.06 | 59.28 |
| Feeder 1 ratio | - | 0.10 | 1.00 | 0.32 | 0.09 |
| Feeder 2 ratio | - | 0.50 | 1.70 | 1.40 | 0.08 |
| Feeder 3 ratio | - | 0.05 | 1.00 | 0.22 | 0.09 |
| Pebble crusher 1 working status | - | 0 | 2 | 0.88 | 0.99 |
| Pebble crusher 2 working status | - | 0 | 2 | 1.02 | 1.00 |
| %PL 13.2mm | % | 5.30 | 99.30 | 26.30 | 3.99 |
| %PL 13.2-19mm | % | 0.00 | 82.70 | 5.19 | 1.43 |
| %PL 19-26.5mm | % | 0.00 | 25.00 | 8.28 | 0.98 |
| %PL 26.5-37.5mm | % | 0.00 | 34.00 | 11.42 | 1.56 |
| %PL 37.5-53mm | % | 0.00 | 16.00 | 10.27 | 0.77 |
| %PL 53-75mm | % | 0.00 | 24.00 | 12.91 | 1.24 |
| %PL 75-106mm | % | 0.00 | 71.00 | 14.74 | 1.93 |
| %PL 106-150mm | % | 0.00 | 35.00 | 8.65 | 2.73 |
| %PL 150-212mm | % | 0.00 | 5.00 | 1.60 | 0.87 |
| %PL 212-300mm | % | 0.00 | 1 | 0.67 | 0.47 |
| P80 | mm | 4.00 | 100.00 | 83.94 | 9.18 |

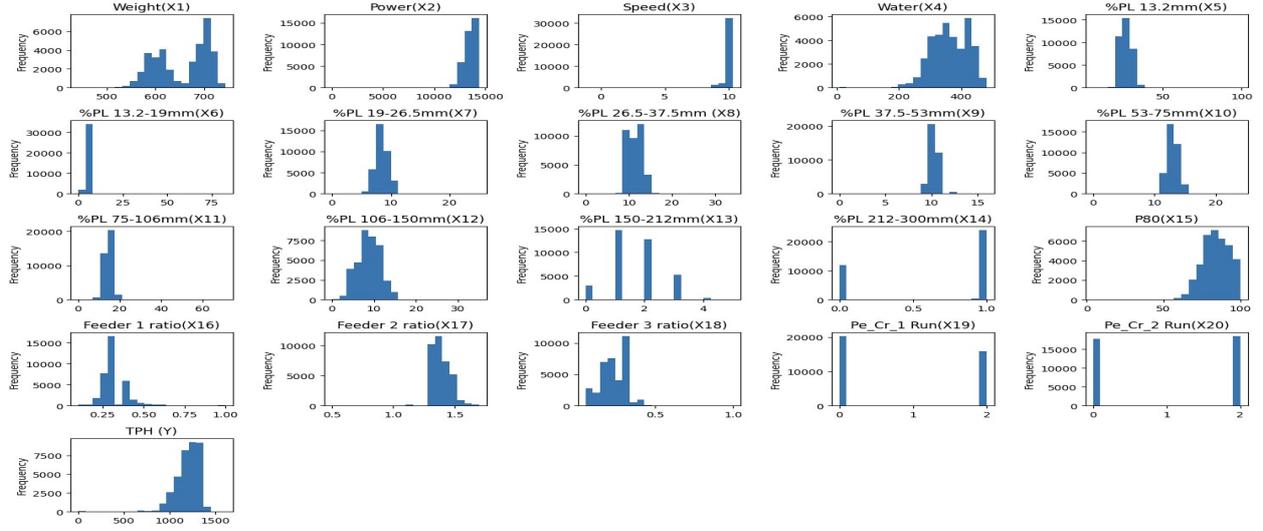

Figure 2: Histogram plot of all features

## 3. Methodology

### 3.1. Outlier detection and feature selection methods

In this study, we utilise local outlier factor (LOF) [9] for outlier detection and recursive feature elimination (RFE) [10] for selecting the most influential features. LOF calculates an outlier factor for each data point based on the local density around it as represented below:

$$LOF_k(p) = \frac{\sum_{o \in N_k(p)} lrd_k(o)/lrd_k(p)}{|N_k(p)|} \qquad (1)$$

where $N_k(p)$ denotes the set of $k$ nearest neighbours of data point $p$, $lrd_k(o)$ is the local reachability density of a neighbour data point $o$, and $lrd_k(p)$ is the local reachability density of the data point $p$. Data points with high *LOF* values are considered outliers, as these values indicate that the local density of that data point is significantly lower than that of its neighbours, suggesting that the data point is in a sparser region of the feature space.

Recursive Feature Elimination (RFE) is a potent feature selection technique. In this approach, the least influential factors are removed in an iterative process. To apply this technique, the prediction method and the desired number of final features need to be determined. Initially, the model is trained using the entire set of features, and ranks are assigned to features based on their impact on predictive accuracy. The least important feature is then removed, and the model is retrained using the remaining features. This process continues until the specified number of features is reached.

### 3.2. Prediction methods

#### 3.2.1. Classic machine learning models

There is a wide range of classic prediction models that can be utilized for predictive modelling. In this study, we employed several well-known machine learning models including linear regression (LR), Lasso regression [11], Elastic Net regression [12], stochastic gradient descent (SGD) [13], support vector machine (SVM)[14], Bayesian regression[15], and K-Nearest Neighbours (KNN) [16]. All of these methods, except for KNN, provide linear regression equations for prediction but employ distinct techniques, cost functions, and optimization strategies. The standard cost function of the linear regression method aims to minimise the sum of squared prediction errors as shown in 2.

$$C_{LR}(w) = \frac{1}{2}\sum_{i=1}^{n}(y_i - w^T x_i)^2 \qquad (2)$$

where the regression model is ˆy = b+w$^T$x, and y, x, b, and w are the target, explanatory, intercept, and slope variables, respectively. Lasso (Least absolute shrinkage and selection operator) and Elastic Net are variations of linear regression that add penalty terms to the linear regression cost function. LASSO adds a penalty term to this objective function, as shown in 3, which is the summation of the absolute values of the variable's coefficients.

$$C_L(w) = \frac{1}{2}\sum_{i=1}^{n}\left(y_i - w^T \phi(x_i)\right)^2 + \lambda \parallel w \parallel_1 \qquad (3)$$

in this equation ϕ is a kernel function and λ is the regularization parameter. Incorporation of this L1 regularization term equips the model with an automatic feature selection property, which leads to a simplified model containing the most influential predictors. Elastic Nets, as indicated by 4, incorporate L2 regularization to the linear regression cost function in addition to L1 regularization, which is the summation of the squared regression coefficient.

$$C_{EN}(w) = \frac{1}{2}\sum_{i=1}^{n}\left(y_i - w^T \phi(x_i)\right)^2 + \lambda_1 \parallel w \parallel_1 + \lambda_2 (w)^2 \qquad (4)$$

SGD regression utilizes the stochastic gradient descent method to iteratively minimize the regression's cost function. It calculates the gradient of the cost function with respect to the model parameters as shown in 5, and adjusts the parameters in the opposite direction of the gradient by utilizing a learning rate α: $w_{t+1} = w_t + \alpha \nabla J(w_t)$.

$$\nabla J(w) = -(y_i - w^T x_i) x_i \qquad (5)$$

In contrast to the conventional linear regression approach, which assumes fixed values for regression coefficients, Bayesian regression offers a more robust framework by considering these coefficients as stochastic variables and modeling their probability distributions. This Bayesian perspective addresses the inherent uncertainty associated with parameter estimation.

Support vector regression (SVR), an extension of support vector machines (SVM) [17], for continuous output variables, aims to find a function *f(x)* such that data points to be positioned within an ε-margin of the regression function. Only those training data points which are placed outside of this margin incur penalties. The ε-SVR can be modeled as the following quadratic programming problem.

$$\min_{w,b,\xi,\xi^*} \frac{1}{2}\|w\|^2 + C\mathbf{e}^T(\xi + \xi^*)$$
$$\text{s.t.} \quad y - (Aw + b\mathbf{e}) \leq \varepsilon\mathbf{e} + \xi, \quad \xi \geq 0$$
$$(Aw + b\mathbf{e}) - y \leq \varepsilon\mathbf{e} + \xi^*, \quad \xi^* \geq 0$$

where $A = [X_1, X_2,..., X_m]^T$ and y are the training data, e is a vector of ones, ξ and ξ* are slack variables penalizing samples outside the ε-margin and C is the regularization factor balancing fitting errors and model simplicity. The ε-SVR can employ kernel functions to handle nonlinear regression problems and discover optimal hyperplanes.

K-Nearest Neighbours (KNN) calculates the distance between a data point and all training data points, identifying k data points with the shortest distances as the k nearest neighbours. The predicted target variable for this data point is the simple or weighted average of the target values of the k-nearest neighbours.

### *3.2.2. Tree-based models*

*Decision trees (DTs).* [18] divide the initial data set into homogeneous subsets with respect to the response variable through developing tree structures. These structures are generated using tree-growing and pruning processes. In the growing step, branches in this tree structure are formed based on extracted rules for input variables using recursive binary partitioning. After forming the initial tree, during the pruning process, insignificant leaves will be removed. Regression models will be fitted to data points assigned to leaf nodes and will be used to produce predictions for new samples. Regression trees are interpretable and flexible models capable of capturing higher-order nonlinear patterns in data.

*Extra trees (ETs).* [19] generate a large number of decision trees, where each tree is constructed using all the training data points. In this approach, a random subset of features is selected at each split point of a decision tree, and the splitting value is also randomly selected. The final prediction is computed by averaging the predictions of all individual decision trees. Compared with traditional DTs, ETs are more robust and less prone to overfitting due to the introduced randomness in feature selection and threshold determination.

*3.2.3. Neural network models*

Two powerful neural network models, including MLP (Multi-Layer Perceptron) [20] and LSTM (Long ShortTerm Memory) [21] are also employed in this study to leverage the power of deep learning. MLP consists of an input layer, one or more hidden layers, and an output layer. The number of nodes in the input and output layers aligns with the number of input and output variables, respectively. The configuration of hidden layers, including the number of layers and nodes within each layer, is determined by the user and defines the network's complexity. Each node generates a signal, which is calculated as

$$y_i = f \sum_{i=1}^{n}(x_i, w_i) \quad (6)$$

where $x_i$, $w_i$, and $f(.)$ are inputs, connection weights, and activation functions, respectively. Throughout the training process, connection weights are iteratively adjusted to enhance the accuracy of output variable prediction. To mitigate the risk of overfitting during training, the early stopping method [22] is used. In this method, a validation set, distinct from the training data, is used to assess the model's performance. The prediction error on this validation set serves as the stopping criterion for training. Backpropagation (BP) [23] is widely used for training MLPs. In this method, each connection weight is updated using a simple gradient descent algorithm as follows.

$$w_{ij}(t+1) = w_{ij}(t) - \eta \frac{\partial E}{\partial w_{ij}(t)} \quad (7)$$

where $w_{ij}(t+1)$ is the updated weight connecting neurons $i$ and $j$ at time $t+1$, $w_{ij}(t)$ is the weight at time $t$, $\eta$ is the learning rate, and $\frac{\partial E}{\partial w_{ij}(t)}$ represents the partial derivative of the error $E$ with respect to $w_{ij}(t)$, indicating the weight update direction and magnitude.

Certain data types, such as biological, text and process data, have sequential dependencies where past values affect current values. Ignoring these dependencies can result in significant information loss [24]. To address this issue, specialized neural networks entitled recurrent neural networks (RNNs) [25] have been developed. The structure of an RNN resembles a typical MLP but includes additional hidden units to capture sequence dependencies. These additional interconnections between hidden units allow the passage of information from one step to the next step, enabling the discovery of temporal correlations. One major challenge with RNNs is their limited ability to capture long-term dependencies in sequences due to difficulties with backpropagation over extended time lags. To address this challenge, Long Short-Term Memory (LSTM) [21] networks were introduced. These networks, as illustrated in Figure 3, incorporate the input, output, and forget gates that selectively control information flow to address the long-term dependency challenge.

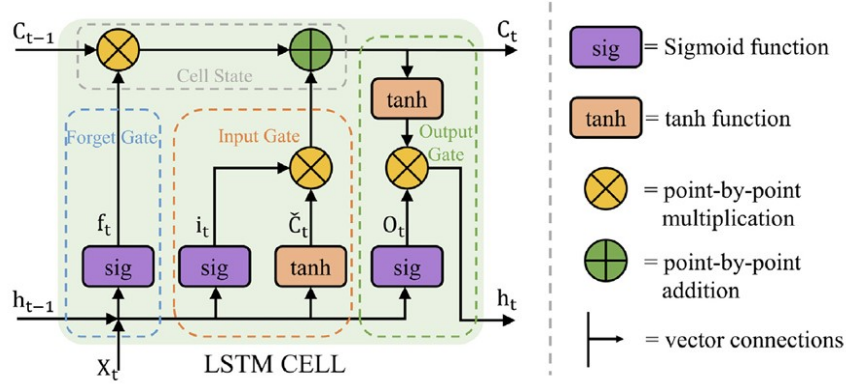

Figure 3: The structure of a basic LSTM cell ([26])

As shown in Figure 3, $X_t$ (the external input) and $h_t-1$ (the upper output) enter the forget gate. The forget gate filters and retains relevant information as shown in (8). The input gate determines what information should be stored in the unit state, as shown in (9). The input is activated using the hyperbolic tangent (tanh) function to obtain output $\tilde{c}_t$, as shown in (10). These three outputs are used to update the memory cell state from the previous state ($c_{t-1}$) to the current state $c_t$, as shown in (11). Finally, the output $o_t$ is obtained through a calculation, as shown in (12), and the final output of the LSTM unit, denoted as $h_t$ or $y_t$, is calculated by multiplying the output $o_t$ with the new state $c_t$ (after activation through tanh), as shown in (13).

$$f_t = \sigma(W_f \cdot [h_{t-1}, x_t] + b_f) \tag{8}$$
$$i_t = \sigma(W_i \cdot [h_{t-1}, x_t] + b_i) \tag{9}$$
$$\tilde{c}_t = \tanh(W_c \cdot [h_{t-1}, x_t] + b_c) \tag{10}$$
$$c_t = f_t \cdot c_{t-1} + i_t \cdot \tilde{c}_t \tag{11}$$
$$o_t = \sigma(W_o \cdot [h_{t-1}, x_t] + b_o) \tag{12}$$
$$y_t = h_t = o_t \cdot \tanh(c_t) \tag{13}$$

*3.2.4. Ensemble models*

In contrast to common data-driven modelling approaches that involve fitting a single predictive model, ensemble based models [27] combine the predictions of several base estimators to enhance robustness. The fundamental ensemble learning methods include bagging [28], boosting [29], and stacking [30]. In bagging, several subsets of training data are randomly selected with replacement and used to train the same model. The training of models is independent of each other and can be performed in parallel. The final prediction is the average of predictions by all models. Bagging improves generalisation and reduces overfitting. In contrast to bagging, the boosting approach involves sequential training of models, where each new model aims to enhance prediction by focusing on weak predictions of previous models in the ensemble. Boosting is one of the most powerful learning approaches introduced in machine learning [31]. Stacking is a more complex ensemble method. Instead of directly averaging or sequentially combining base model predictions, like bagging and boosting, stacking uses these individual base models' predictions as input for a higher-level model. This meta-model is trained on these diverse base model predictions to make accurate final predictions.

In this study, we employ random forest (RF) regression [32] as a bagging method and utilize several boosting methods including AdaBoost (Adaptive Boosting) [29], GBM (Gradient Boosting Machine) [33], XGBoost (Extreme Gradient Boosting) [34], CatBoost (Categorical Boosting) [35], and HGBM (Histogram-based Gradient Boosting Machine) [36] as prediction models.

*Random Forest (RF).* This regression method involves creating a large number of diverse RTs to reduce overfitting and improve generalization. Each of the RTs is generated using a random subset of training samples and a randomly selected subset of variables at each node. To obtain final predictions, the average value of predictions from all RTs is calculated. This approach not only enhances accuracy and robustness in regression tasks but also equips the model to effectively handle high-dimensional data [37].

*Adaptive Boosting (AdaBoost).* This boosting method takes advantage of sequential learning by assigning and updating weights to training samples based on prediction errors in a progressive manner. It starts with assigning the same weight to all training samples. In each iteration, AdaBoost selects a random subset of the training data based on the current weights, trains a weak learner to minimize prediction errors, adjusts the data point weights by assigning larger weights to data points that were previously poorly predicted, and repeats the process until certain termination criteria are met which could be reaching the maximum number of generated models or achieving a specified quality threshold in the predictions. The final regression prediction is a weighted combination of the predictions from all the weak learners. This focus on challenging examples guides the training to become specialized in areas where the ensemble previously worked weakly.

*Gradient Boosting Machine (GBM).* This method follows a forward stage-wise strategy by sequentially adding new predictive models to the ensemble. It commonly employs regression trees (RTs) as weak learners. These RTs are generated consecutively, attempting to model the prediction residuals from previous RTs. GBM begins by defining a differentiable loss function, often using the squared error as the loss function, which is expressed as:

$$L(y, F(x)) = \tfrac{1}{2}(y - F(x))^2 \qquad (14)$$

where $y$ and $F$ represent real output values and weak learner's predictions, respectively. The initialization step involves considering a constant prediction value for all data points, computed as:

$$F_0(x) = \text{argmin}_\gamma \sum_{i=1}^n L(y_i, \gamma) \qquad (15)$$

If 14 is used as the loss function, the results of 15 are the simple average of real outputs for training data. It then computes the pseudo-residuals for each data point as:

$$r_{im} = -\left[\frac{\partial L(y_i, F(x_i))}{\partial F(x_i)}\right]_{F(x)=F_{m-1}(x)}, \text{ for } i = 1, \dots, n \qquad (16)$$

Subsequently, GBM fits a base learner (e.g., a tree) scaled to the pseudo-residuals, denoted as $h_m(x)$, using the training set $\{(x_i, r_{im})\}^n_{i=1}$. In the following step, GBM computes the multiplier $\rho_m$ by solving the optimization problem:

$$\rho_m = \underset{\rho}{\text{argmin}} \sum_{i=1}^n L(y_i, F_{m-1}(x_i) + \rho h_m(x_i)) \qquad (17)$$

and finally updates the model as below:

$$F_m(x) = F_{m-1}(x) + \gamma_m, \text{ where } \gamma_m = \rho_m h_m(x) \qquad (18)$$

*Extreme Gradient Boosting (XGBoost).* This method improves GBM by incorporating regularization techniques to avoid overfitting and complex modelling. The objective function of this state-of-the-art method, as represented in 19, consists of both training loss and model complexity to develop simple prediction models that have strong predictive performance.

$$\text{Objective}(F) = \sum_{i=1}^{n}\left[l(y_i, F(x_i)) + \Omega(f_i)\right] \quad (19)$$

In this function, the first section $l(y_i, F(x_i))$ is the loss function calculating the difference between real and predicted values of the output variable, and $\Omega(f_i)$ is the regularization penalty term that encourages a simpler model. It commonly includes both L1 and L2 regularisations as below:

$$\Omega(f_t) = \gamma T + \frac{1}{2}\alpha \sum_{j=1}^{T}|w_j| + \frac{1}{2}\lambda \sum_{j=1}^{T} w_j^2 \quad (20)$$

In this equation, there are three regularisation parameters, including γ, α, and λ. γ multiplied by $T$ as the number of leaf nodes controls the model complexity. α, and λ are $L1$ and $L2$ regularization term penalty coefficients, respectively.

*Histogram-based Gradient Boosting Machine (HGBM).* This variant of gradient boosting is highly efficient for modelling large data sets. The methodology involves creating histograms for each continuous feature within the dataset. This involves discretising the data into bins, reducing the number of potential segmentation points to be considered. Subsequently, the algorithm identifies optimal segmentation points by analysing these histogram bins. Furthermore, utilizing histograms instead of continuous variables has a positive impact on memory usage, particularly for large datasets. This memory-efficient approach not only enhances the algorithm's capacity to handle extensive data but also ensures efficient computational speed and high accuracy.

*Categorical Boosting (CatBoost).* It is another gradient boosting method that can effectively handle both categorical and numerical features without any feature encoding. In this method, as illustrated in Figure 4, an ordered boosting technique is applied to avoid the target leakage problem. In this technique, first n random samples are selected from the training set and used to train prediction models $M_1, M_2, ..., M_{n-1}$ in such a way that for training each model $M_i$, $i$ samples from the selected training data points are used.

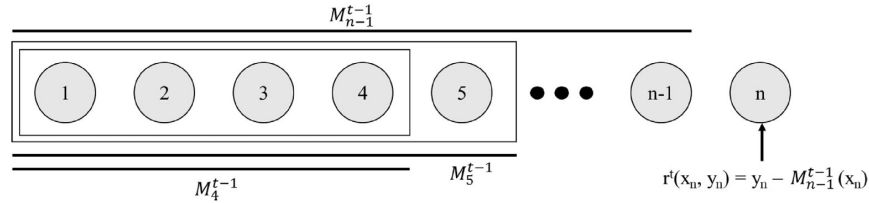

Figure 4: Ordered boosting principle [35]

In contrast to standard gradient boosting, where residuals are computed using the same dataset, ordered boosting addresses the issue of data leakage and overfitting by determining the residual $r^t$ for the $j$th sample in the $t$th iteration based on the prediction made by $M_{j-1}$. The gradient is then updated using the $t$th residual value to construct the $t$th model.

### 3.3. Evolutionary algorithms for optimisation
### 3.3.1. Differential Evolution (DE) algorithm

Differential Evolution (DE) [38] is an efficient population-based optimization algorithm. This method is preferred in many practical optimization problems due to its ease of use, consistent stability, minimal parameter demands, and robustness. DE is comprised of four steps, including population initialization, mutation, crossover, and selection. First, a random generation of initial solutions is created as presented below:

$$x_i = x_{\min,j} + \text{rand}(x_{\max,j} - x_{\min,j}) \quad (21)$$

where $x_{\min,j}$ and $x_{\max,j}$ represent lower and upper limits of the *j*th decision variable, respectively. Various mutation strategies are applicable in the context of DE. These strategies vary in their global search capabilities and convergence speed. While some strategies emphasize robust global exploration through random individual selection, others prioritize faster convergence by incorporating optimal vector information. The most classic mutation strategy involves updating a solution's position by computing the difference between two vectors with respect to the target vector as expressed as follows:

$$V_i^t = X_i^t + F \cdot (X_{r1}^t - X_{r2}^t) \tag{22}$$

In this equation, $X_i^t$, $X_{r1}^t$, and $X_{r2}^t$ represent distinct individuals, and $i$, $r1$, $r2$. $X_i^t$ is the individual undergoing mutation, referred to as the target vector, while $X_{r1}^t$ and $X_{r2}^t$ are individuals chosen for the difference computation. The result, $V_i^t$, represents the mutated vector derived from $X_i^t$ and is known as the mutation vector. The parameter $F$ corresponds to the scaling factor. During the crossover step, the mutation vector $V_i^t$ and the target vector $X_i^t$ are recombined to generate the trial vector $U_i^t$ to enhance population diversity. Finally, in the selection step, the fitness values of trial and target vectors are compared and the individual with the better target value is selected as the target vector of the next generation.

### 3.3.2. Genetic Algorithms (GAs)

Genetic algorithms (GAs) [39] are population-based optimization methods inspired by natural evolution and are widely employed to tackle complex optimization problems. In a GA, individuals are represented as chromosomes, where each chromosome consists of genes that correspond to the variables targeted for optimization. The GA algorithm encompasses several essential steps.

Firstly, the algorithm begins by generating an initial population of chromosomes as individuals. In the next step, the fitness of each individual within the population is assessed using a dedicated fitness function. This function quantifies how well a particular solution aligns with the optimization objective, guiding the search for better solutions. Selection is a critical step in GAs, where individuals from the current population are chosen as parents for the next generation. The probability of selection is typically determined by an individual's fitness. There are various selection methods, such as uniform selection, elitist selection, and tournament selection. Selected parents undergo genetic operations to create new individuals. Crossover, or recombination, combines genes from two parents to create offspring. Additionally, mutation introduces random alterations to individual chromosomes, ensuring diversity within the population and preventing premature convergence.

The newly generated offspring collectively form the updated generation. Each individual is evaluated again using the fitness function, and the best-performing individuals are selected to become parents for the succeeding generation. Finally, the GA process continues to iterate through multiple generations until it satisfies the termination criterion, which can be reaching a maximum generation limit or achieving a predefined quality threshold.

### 3.3.3. Particle Swarm Optimisation (PSO)

PSO [40, 41] is one of the most popular swarm-based optimisation methods utilizing the collective intelligence of a swarm of particles to efficiently explore a complex search space and find optimal solutions. In the PSO algorithm, a population of particles or candidate solutions travels through a multi-dimensional search space to find the optimal solution. Each particle adjusts its position and velocity based on its own and its neighbour's experience. This dynamic adjustment is guided by the particle's best prior position and the global best position achieved by the entire population. The PSO algorithm in a D-dimensional search space updates the position vector of the *i*th particle $x_i = (x_{i1}, x_{i2},..., x_{iD})$ and its velocity $v_i = (v_{i1}, v_{i2},..., v_{iD})$ using equations 23 and 24, respectively.

$$x_{i,d}^{t+1} = x_{i,d}^t + v_{i,d}^{t+1} \tag{23}$$

$$v_{i,d}^{t+1} = wv_{i,d}^t + c_1r_1(P_{i,d}^t - x_{i,d}^t) + c_2r_2(GP_{i,d}^t - x_{i,d}^t) \quad (24)$$

where $x_{i,d}^{t+1}$ and $v_{i,d}^{t+1}$ represent the position and velocity of particle *i* in dimension *d* at the next iteration, respectively. $P_{i,d}^t$ is the optimal historical position of the particle *i* in dimension *j*, and $GP_{i,d}^t$ is the overall best position achieved by the entire swarm up to that iteration. The parameter *w* is inertia weight. $c_1$, and $c_2$ are cognitive and social acceleration coefficients, adjusting exploitation and exploration of the search space, respectively. $r_1$ and $r_2$ are random values with a uniform distribution between 0 and 1.

*3.4. The proposed framework*

Figure 5 illustrates our proposed framework for maximising SAG mill throughput. As depicted in this figure, after receiving the initial data set, in collaboration with an industry expert, relevant features are selected from an extended list of potential features. The data set undergoes a cleaning process to eliminate erroneous data. Subsequently, this refined dataset is employed to train 17 machine learning (ML) prediction models. Prediction results are compared, and the best model is identified. Feature selection and outlier detection methods are applied to increase the model's prediction accuracy. The best model, trained with the final selected features, is then integrated into three distinct evolutionary algorithms as the fitness function and the performance of optimisation algorithms are compared, considering lower and upper bounds of input features as constraints. The integrated framework, including the best prediction model and the best optimisation algorithm, is able to propose optimal parameter settings that can lead to maximum mill throughput. It is noteworthy that default parameters are used for all ML and AI models.

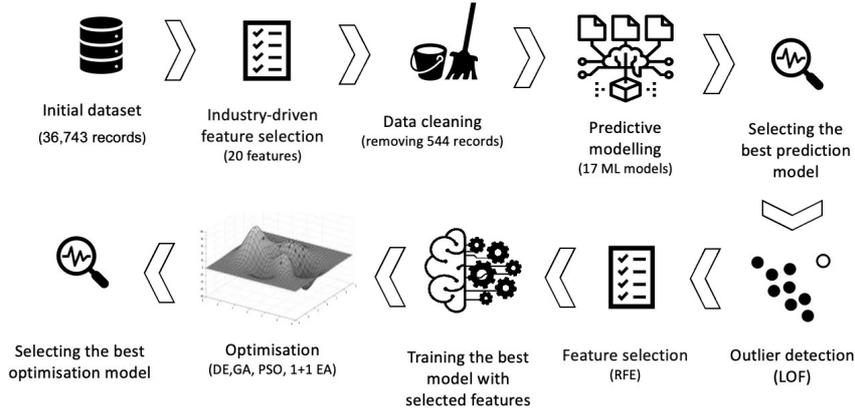

Figure 5: The proposed intelligent framework for maximising SAG mill throughput

## 4. Experimental results

This study unveils the outcomes of SAG mill throughput maximisation attained through a proposed novel and efficient hybrid surrogate model, employing a comprehensive machine learning framework integrated with the top performing optimiser. In the initial phase, we conducted a rigorous evaluation of 17 renowned ML models, encompassing a fusion of linear-based regressions, SVM, Neural networks, Decision tree-based approaches, and six distinct types of Ensemble models. Upon evaluating the metrics outlined in Table 2, we identified the most proficient ML model, which we subsequently employed as a precise and expeditious surrogate model to estimate SAG mill throughput. Moving forward, we crafted an optimisation framework incorporating three widely used techniques: GA, PSO, and DE. The performance of these optimisation methods was meticulously assessed, taking into account their convergence rate and the calibre of solutions generated. Our comprehensive analysis of the results and subsequent discussions yielded valuable insights and recommendations.

*4.1. Machine learning prediction results*

In this study, we comprehensively evaluated ML frameworks encompassing a diverse set of 17 models. This evaluation entailed employing an extensive array of metrics (See Table 2). In these metrics, $N_s$ represents the total number of samples, $k$ is the index of each individual sample, $f_e(k)$ is the predicted mill throughput for the $k$-th sample, and $f_t(k)$ is its real value. These metrics facilitated an objective comparison of various models or iterations of the same model, thereby quantifying their precision and predictive power. Furthermore, utilizing these evaluation metrics proved instrumental in carefully selecting the optimal model for a given task, allowing for considering specific requirements and priorities. Through an accurate analysis of these metric values, researchers and practitioners alike could discern areas necessitating model refinement and optimisation. Ultimately, the incorporation of these evaluation metrics enabled a systematic and quantitative assessment of the models' performance, ensuring an accurate and objective appraisal.

Table 2: The performance evaluation metrics for the prediction models

| Metrics | Definition | Equation |
| --- | --- | --- |
| MSE | Mean Square Error | $MSE = \frac{1}{N_s} \sum_{k=1}^{N_s} (f_e(k) - f_t(k))^2$ |
| RMSE | Root Mean Square Error | $RMSE = \sqrt{\frac{1}{N_s} \sum_{k=1}^{N_s} (f_e(k) - f_t(k))^2}$ |
| MAE | Mean Absolute Error | $MAE = \frac{1}{N_s} \sum_{k=1}^{N_s} |f_e(k) - f_t(k)|$ |
| MAPE | Mean Absolute Percentage Error | $MAPE = \frac{1}{N_s} \sum_{t=1}^{N_s} \left| \frac{f_e(k) - f_t(k)}{f_t(k)} \right|$ |
| SMAPE | Symmetric Mean Absolute Percentage Error | $SMAPE = \frac{1}{N_s} \sum_{k=1}^{N_s} \frac{|f_t(k) - f_e(k)|}{(\frac{1}{2}(f_t(k) + f_e(k)))} \times 100$ |
| R-value | Pearson Correlation Coefficient | $R = \frac{\frac{1}{N_s} \sum_{k=1}^{N_s} (f_e(k) - \bar{f_e})(f_t(k) - \bar{f_t})}{\sqrt{\frac{1}{N_s} \sum_{k=1}^{N_s} (f_e(k) - \bar{f_e})^2} \times \sqrt{\frac{1}{N_s} \sum_{k=1}^{N_s} (f_t(k) - \bar{f_t})^2}}$ |
| EVS | Explained Variance Score | $EVS = 1 - \frac{\sum_{k=1}^{N_s} Variance(f_t(k) - f_e(k))}{Variance(f_t(k))}$ |
| MSLE | Mean Squared Logarithmic Error | $MSLE = \frac{1}{N_s} \sum_{k=1}^{N_s} (\log_e(1 + f_t(k)) - \log_e(1 + f_e(k)))^2$ |

K-fold cross-validation with k=10 is used to compare prediction methods, and the prediction performance of the employed models based on evaluation metrics are presented in Tables 3, 4, and 5.

The statistical performance outcomes of six supervised learning ML models, namely Support Vector Machines (SVM), Linear Regression, LASSO (Least Absolute Shrinkage and Selection Operator), Stochastic Gradient Descent (SGD), Elastic Net, and Bayesian models, are presented in Table 3. These models exhibit the capability to yield interpretable models by directly estimating coefficients, thereby enabling valuable insights regarding the significance and orientation of features.

As can be seen in Table 3, among the six models introduced, Elastic Net demonstrated the best performance based on the evaluation metrics and criteria employed in the study. There could be several reasons why Elastic Net outperformed the other models. First and foremost, Elastic Net combines both L1 (LASSO) and L2 (ridge regression) regularization techniques. This hybrid approach allows for effective feature selection by shrinking less significant coefficients towards zero (L1 regularization) while also handling multicollinearity among features (L2 regularization). This flexibility in regularization can lead to improved model performance. Furthermore, Elastic Net is known for its robustness in dealing with outliers in the data. It achieves this by utilizing both L1 and L2 regularization, which helps downweigh the outliers' impact on the model's coefficients.

Table 3: Statistical prediction results of grinding mill throughput using six well-known machine learning models with seven evaluation metrics

| | SVM | | | | | | | | Linear Regression (LR) | | | | | | |
| --- | --- | --- | --- | --- | --- | --- | --- | --- | --- | --- | --- | --- | --- | --- | --- |
| | RMSE | MAE | EVS | MSLE | SMAPE | MAPE | R2 | | RMSE | MAE | EVS | MSLE | SMAPE | MAPE | R2 |
| Min | 5.20E+01 | 4.35E+01 | -5.25E-01 | 1.56E-03 | 3.30E+00 | 3.23E+00 | 5.39E-02 | Min | 5.34E+01 | 3.87E+01 | -2.82E+00 | 1.81E-03 | 3.12E+00 | 3.07E+00 | 6.50E-02 |
| Max | 1.15E+02 | 7.75E+01 | 4.19E-01 | 1.91E-01 | 6.65E+00 | 1.13E+03 | 5.19E-01 | Max | 9.26E+01 | 7.32E+01 | 5.82E-01 | 2.54E-01 | 6.03E+00 | 6.01E+02 | 6.06E-01 |
| Mean | 7.98E+01 | 6.11E+01 | 2.19E-01 | 2.92E-02 | 5.04E+00 | 1.49E+02 | 3.41E-01 | Mean | 7.32E+01 | 5.68E+01 | 1.06E-01 | 5.10E-02 | 4.68E+00 | 8.18E+01 | 4.39E-01 |
| Median | 7.68E+01 | 5.71E+01 | 2.89E-01 | 4.30E-03 | 4.61E+00 | 4.70E+00 | 3.82E-01 | Median | 7.24E+01 | 5.96E+01 | 4.66E-01 | 3.50E-03 | 4.77E+00 | 4.72E+00 | 5.00E-01 |
| STD | 1.83E+01 | 1.06E+01 | 2.60E-01 | 5.60E-02 | 1.10E+00 | 3.34E+02 | 1.47E-01 | STD | 1.35E+01 | 1.05E+01 | 9.91E-01 | 9.12E-02 | 9.31E-01 | 1.78E+02 | 1.69E-01 |
| | LASSO | | | | | | | | SGD | | | | | | |
| | RMSE | MAE | EVS | MSLE | SMAPE | MAPE | R2 | | RMSE | MAE | EVS | MSLE | SMAPE | MAPE | R2 |
| Min | 4.68E+01 | 3.94E+01 | -2.82E+00 | 1.30E-03 | 3.04E+00 | 3.00E+00 | 6.48E-02 | Min | 4.72E+01 | 3.57E+01 | -4.49E+00 | 1.41E-03 | 2.87E+00 | 2.86E+00 | 1.69E-02 |
| Max | 9.20E+01 | 7.21E+01 | 6.11E-01 | 2.57E-01 | 5.98E+00 | 6.46E+02 | 6.13E-01 | Max | 1.02E+02 | 8.04E+01 | 6.45E-01 | 3.39E-01 | 6.85E+00 | 7.21E+02 | 6.76E-01 |
| Mean | 7.22E+01 | 5.55E+01 | 7.44E-02 | 4.66E-02 | 4.58E+00 | 8.33E+01 | 4.20E-01 | Mean | 7.48E+01 | 5.82E+01 | 6.63E-02 | 4.93E-02 | 4.79E+00 | 8.07E+01 | 4.21E-01 |
| Median | 7.49E+01 | 5.75E+01 | 3.94E-01 | 3.71E-03 | 4.66E+00 | 4.60E+00 | 4.73E-01 | Median | 7.25E+01 | 5.90E+01 | 4.31E-01 | 3.60E-03 | 4.70E+00 | 4.66E+00 | 4.76E-01 |
| STD | 1.57E+01 | 1.14E+01 | 9.83E-01 | 8.49E-02 | 1.05E+00 | 1.91E+02 | 1.81E-01 | STD | 1.44E+01 | 1.14E+01 | 1.04E+00 | 9.17E-02 | 9.98E-01 | 1.82E+02 | 1.68E-01 |

|        | Elastic Net |          |           |          |          |          |          |        | Bayesian |          |           |          |          |          |          |
|--------|-------------|----------|-----------|----------|----------|----------|----------|--------|----------|----------|-----------|----------|----------|----------|----------|
|        | RMSE        | MAE      | EVS       | MSLE     | SMAPE    | MAPE     | R2       |        | RMSE     | MAE      | EVS       | MSLE     | SMAPE    | MAPE     | R2       |
| Min    | 8.71E+01    | 5.91E+01 | -2.92E-03 | 4.56E-03 | 5.59E+00 | 5.75E+00 | 2.83E-04 | Min    | 5.33E+01 | 3.88E+01 | -2.82E+00 | 1.81E-03 | 3.12E+00 | 3.08E+00 | 6.47E-02 |
| Max    | 1.34E+02    | 1.09E+02 | 5.31E-02  | 1.90E-01 | 9.14E+00 | 1.12E+03 | 1.59E-01 | Max    | 9.25E+01 | 7.31E+01 | 5.83E-01  | 2.55E-01 | 6.02E+00 | 5.95E+02 | 6.05E-01 |
| Mean   | 1.08E+02    | 8.78E+01 | 2.03E-02  | 3.25E-02 | 7.20E+00 | 1.49E+02 | 5.75E-02 | Mean   | 7.32E+01 | 5.68E+01 | 1.05E-01  | 5.01E-02 | 4.68E+00 | 8.10E+01 | 4.38E-01 |
| Median | 1.06E+02    | 8.50E+01 | 1.76E-02  | 7.91E-03 | 6.82E+00 | 7.35E+00 | 4.33E-02 | Median | 7.23E+01 | 5.96E+01 | 4.67E-01  | 3.50E-03 | 4.76E+00 | 4.71E+00 | 4.99E-01 |
| STD    | 1.60E+01    | 1.72E+01 | 1.67E-02  | 5.47E-02 | 1.33E+00 | 3.31E+02 | 5.26E-02 | STD    | 1.35E+01 | 1.04E+01 | 9.91E-01  | 8.97E-02 | 9.29E-01 | 1.76E+02 | 1.69E-01 |

Table 4 shows the prediction mill throughput results of four popular ML models: Decision tree (DT), Extra Tree (ET), MLP and a recurrent neural model (LSTM). The highest prediction accuracy with minimum RMSE was provided by Extra Tree (ET). While Extra Trees (ET) bear resemblances to Decision Trees (DT), they distinguish themselves by incorporating an extra layer of randomness into the tree construction process. This infusion of randomness imparts an augmented level of diversity, effectively mitigating the risk of overfitting by reducing variance. The unique attribute of ETs lies in their exceptional ability to navigate the complex topography of high-dimensional datasets characterized by noise. They exhibit a remarkable aptitude for dealing with the intricacies presented by such datasets, enabling them to deliver robust performance across diverse domains.

Table 4: Statistical prediction results of grinding mill throughput using two popular Tree-based models (DT and ET) and two Neural models (MLP and LSTM) with seven evaluation metrics

|        | Decision Tree (DT) |          |           |          |          |          |          |        | Extra Tree (ET) |          |           |          |          |          |          |
|--------|--------------------|----------|-----------|----------|----------|----------|----------|--------|-----------------|----------|-----------|----------|----------|----------|----------|
|        | RMSE               | MAE      | EVS       | MSLE     | SMAPE    | MAPE     | R2       |        | RMSE            | MAE      | EVS       | MSLE     | SMAPE    | MAPE     | R2       |
| Min    | 2.30E+01           | 1.73E+01 | -1.16E-02 | 3.02E-04 | 1.30E+00 | 1.30E+00 | 2.85E-01 | Min    | 2.59E+01        | 2.02E+01 | -5.74E-01 | 3.76E-04 | 1.52E+00 | 1.51E+00 | 6.06E-02 |
| Max    | 9.20E+01           | 7.04E+01 | 7.04E-01  | 3.39E-02 | 5.95E+00 | 1.34E+02 | 7.74E-01 | Max    | 1.31E+02        | 9.34E+01 | 7.64E-01  | 1.38E-01 | 8.04E+00 | 4.23E+02 | 7.88E-01 |
| Mean   | 6.56E+01           | 4.73E+01 | 3.60E-01  | 8.99E-03 | 3.99E+00 | 1.80E+01 | 4.55E-01 | Mean   | 5.73E+01        | 4.12E+01 | 4.82E-01  | 1.71E-02 | 3.50E+00 | 3.63E+01 | 5.41E-01 |
| Median | 6.54E+01           | 4.82E+01 | 3.70E-01  | 3.10E-03 | 3.96E+00 | 3.96E+00 | 4.31E-01 | Median | 5.72E+01        | 4.12E+01 | 5.63E-01  | 2.45E-03 | 3.44E+00 | 3.51E+00 | 5.80E-01 |
| STD    | 1.93E+01           | 1.47E+01 | 1.84E-01  | 1.13E-02 | 1.37E+00 | 3.91E+01 | 1.25E-01 | STD    | 1.73E+01        | 1.28E+01 | 2.52E-01  | 3.49E-02 | 1.22E+00 | 8.56E+01 | 1.70E-01 |
|        | MLP                |          |           |          |          |          |          |        | LSTM            |          |           |          |          |          |          |
|        | RMSE               | MAE      | EVS       | MSLE     | SMAPE    | MAPE     | R2       |        | RMSE            | MAE      | EVS       | MSLE     | SMAPE    | MAPE     | R2       |
| Min    | 4.05E+01           | 3.29E+01 | -1.58E+00 | 9.23E-04 | 2.47E+00 | 2.46E+00 | 5.94E-02 | Min    | 3.60E+01        | 2.85E+01 | -4.77E+00 | 7.39E-04 | 2.15E+00 | 2.12E+00 | 2.92E-02 |
| Max    | 9.14E+01           | 7.14E+01 | 5.75E-01  | 1.47E-01 | 6.06E+00 | 4.84E+02 | 5.98E-01 | Max    | 2.60E+02        | 1.96E+02 | 6.68E-01  | 1.37E-01 | 1.79E+01 | 4.85E+02 | 7.27E-01 |
| Mean   | 6.91E+01           | 5.21E+01 | 2.52E-01  | 2.18E-02 | 4.35E+00 | 6.15E+01 | 4.30E-01 | Mean   | 8.41E+01        | 6.27E+01 | 6.64E-02  | 2.39E-02 | 5.46E+00 | 4.92E+01 | 5.06E-01 |
| Median | 7.17E+01           | 4.97E+01 | 4.44E-01  | 3.80E-03 | 4.53E+00 | 4.73E+00 | 4.67E-01 | Median | 6.94E+01        | 4.82E+01 | 4.75E-01  | 4.65E-03 | 4.49E+00 | 4.92E+00 | 5.51E-01 |
| STD    | 1.59E+01           | 1.24E+01 | 5.78E-01  | 4.27E-02 | 1.15E+00 | 1.41E+02 | 1.44E-01 | STD    | 4.96E+01        | 3.91E+01 | 1.21E+00  | 3.97E-02 | 3.56E+00 | 1.16E+02 | 2.10E-01 |

As depicted in Table 5, a comparative analysis was conducted to assess the efficacy of six renowned ensemble methods in predicting the SAG mill throughput. The principal rationale behind employing these models stems from the fact that ensemble methods possess the inherent capability to surpass the performance of solitary models through the combination of predictions derived from multiple simple models. By consolidating the heterogeneous knowledge and discernment gleaned from disparate models, the ensemble framework engenders predictions that are not only more precise but also fortified against adversities. This advantage becomes particularly pronounced when the constituent models exhibit complementary strengths and weaknesses. According to the findings presented in Table 5, the ensemble model that exhibited the most impressive performance was CatBoost, as indicated by its minimal RMSE of $4.96E + 01$ and maximal accuracy measured by $R^2$ of $6.63E − 01$ on average. This exceptional superiority can be attributed to the implementation of a diverse range of techniques by CatBoost, specifically devised to tackle the issue of overfitting. Notably, employing a ground-breaking algorithm known as "Ordered Boosting" plays a pivotal role in mitigating overfitting tendencies. By introducing an element of randomness during the tree construction process, this technique effectively enhances the model's capability to generalize well to previously unseen data. Consequently, CatBoost stands out among other ensemble methods, exhibiting reduced susceptibility to overfitting and delivering superior predictive performance.

Table 5: Statistical details of predicting grinding mill throughput using four popular ensemble models with seven evaluation metrics

| | AdaBoost | | | | | | | | GBM | | | | | | |
|---|---|---|---|---|---|---|---|---|---|---|---|---|---|---|---|
| | RMSE | MAE | EVS | MSLE | SMAPE | MAPE | R2 | | RMSE | MAE | EVS | MSLE | SMAPE | MAPE | R2 |
| Min | 6.64E+01 | 5.05E+01 | -1.46E+00 | 3.04E-03 | 4.24E+00 | 4.27E+00 | 1.45E-08 | Min | 2.25E+01 | 1.75E+01 | 2.94E-01 | 2.85E-04 | 1.31E+00 | 1.31E+00 | 3.66E-01 |
| Max | 1.31E+02 | 1.25E+02 | 6.72E-01 | 1.34E-01 | 9.80E+00 | 4.06E+02 | 8.08E-01 | Max | 8.14E+01 | 6.06E+01 | 7.95E-01 | 7.93E-02 | 6.06E+00 | 1.05E+02 | 8.41E-01 |
| Mean | 9.04E+01 | 7.68E+01 | 2.43E-01 | 1.46E-02 | 6.34E+00 | 1.95E+01 | 3.11E-01 | Mean | 5.13E+01 | 3.72E+01 | 6.12E-01 | 1.26E-02 | 3.18E+00 | 1.94E+01 | 6.45E-01 |
| Median | 9.01E+01 | 7.64E+01 | 2.58E-01 | 6.20E-03 | 6.34E+00 | 6.72E+00 | 2.76E-01 | Median | 4.94E+01 | 3.49E+01 | 6.32E-01 | 1.88E-03 | 2.89E+00 | 2.87E+00 | 6.36E-01 |
| STD | 9.63E+00 | 1.24E+01 | 2.50E-01 | 1.79E-02 | 8.37E-01 | 2.56E+01 | 1.99E-01 | STD | 1.56E+01 | 1.23E+01 | 1.29E-01 | 2.24E-02 | 1.30E+00 | 3.04E+01 | 1.20E-01 |
| | HGBM | | | | | | | | XGBoost | | | | | | |
| | RMSE | MAE | EVS | MSLE | SMAPE | MAPE | R2 | | RMSE | MAE | EVS | MSLE | SMAPE | MAPE | R2 |
| Min | 2.23E+01 | 1.71E+01 | 2.91E-01 | 2.79E-04 | 1.28E+00 | 1.28E+00 | 3.38E-01 | Min | 2.35E+01 | 1.77E+01 | 2.91E-01 | 3.13E-04 | 1.33E+00 | 1.32E+00 | 3.82E-01 |
| Max | 1.02E+02 | 7.34E+01 | 7.47E-01 | 1.05E-01 | 7.03E+00 | 1.04E+02 | 7.66E-01 | Max | 8.35E+01 | 6.33E+01 | 7.43E-01 | 3.21E-02 | 5.79E+00 | 8.78E+01 | 8.44E-01 |
| Mean | 5.48E+01 | 3.97E+01 | 5.77E-01 | 1.73E-02 | 3.38E+00 | 2.01E+01 | 6.15E-01 | Mean | 5.84E+01 | 4.23E+01 | 5.12E-01 | 8.80E-03 | 3.58E+00 | 1.47E+01 | 5.63E-01 |
| Median | 4.97E+01 | 3.59E+01 | 6.24E-01 | 1.90E-03 | 3.01E+00 | 3.03E+00 | 6.37E-01 | Median | 5.52E+01 | 3.87E+01 | 5.06E-01 | 2.39E-03 | 3.17E+00 | 3.17E+00 | 5.48E-01 |
| STD | 2.09E+01 | 1.58E+01 | 1.63E-01 | 3.13E-02 | 1.57E+00 | 3.20E+01 | 1.39E-01 | STD | 1.86E+01 | 1.42E+01 | 1.22E-01 | 1.16E-02 | 1.38E+00 | 2.55E+01 | 1.23E-01 |
| | CatBoost | | | | | | | | Random Forest (RF) | | | | | | |
| | RMSE | MAE | EVS | MSLE | SMAPE | MAPE | R2 | | RMSE | MAE | EVS | MSLE | SMAPE | MAPE | R2 |
| Min | 2.22E+01 | 1.73E+01 | 3.00E-01 | 2.76E-04 | 1.29E+00 | 1.29E+00 | 3.53E-01 | Min | 2.32E+01 | 1.81E+01 | 1.81E-01 | 3.03E-04 | 1.36E+00 | 1.35E+00 | 2.92E-01 |
| Max | 7.94E+01 | 5.64E+01 | 7.92E-01 | 1.30E-01 | 5.52E+00 | 3.34E+02 | 8.01E-01 | Max | 9.75E+01 | 6.58E+01 | 8.03E-01 | 5.98E-02 | 6.49E+00 | 5.63E+01 | 8.31E-01 |
| Mean | 4.96E+01 | 3.63E+01 | 6.45E-01 | 1.86E-02 | 3.09E+00 | 4.67E+01 | 6.63E-01 | Mean | 5.53E+01 | 3.91E+01 | 5.69E-01 | 9.62E-03 | 3.34E+00 | 1.10E+01 | 6.15E-01 |
| Median | 4.75E+01 | 3.39E+01 | 6.81E-01 | 1.53E-03 | 2.91E+00 | 2.95E+00 | 7.02E-01 | Median | 4.70E+01 | 3.33E+01 | 6.35E-01 | 1.70E-03 | 2.70E+00 | 2.70E+00 | 6.38E-01 |
| STD | 1.54E+01 | 1.15E+01 | 1.33E-01 | 3.83E-02 | 1.18E+00 | 9.88E+01 | 1.26E-01 | STD | 1.96E+01 | 1.50E+01 | 1.74E-01 | 1.43E-02 | 1.48E+00 | 1.54E+01 | 1.45E-01 |

Figure 6, depicted as a boxplot, concisely presents a comprehensive overview of the distribution of 17 model performance metrics, specifically focusing on the R2 score in the context of SAG mill throughput prediction. This plot effectively captures crucial summary statistics, including the median, quartiles, and potential outliers, facilitating a rapid comprehension of the central tendency and dispersion of the results. Notably, Catboost emerges as the frontrunner regarding accuracy prediction, as evidenced by its superior $R^2$ median and minimal variance across ten independent runs. This outstanding performance positions Catboost as the most reliable and consistent model for accurately predicting SAG mill throughput, as denoted by its robust and impressive $R^2$ scores.

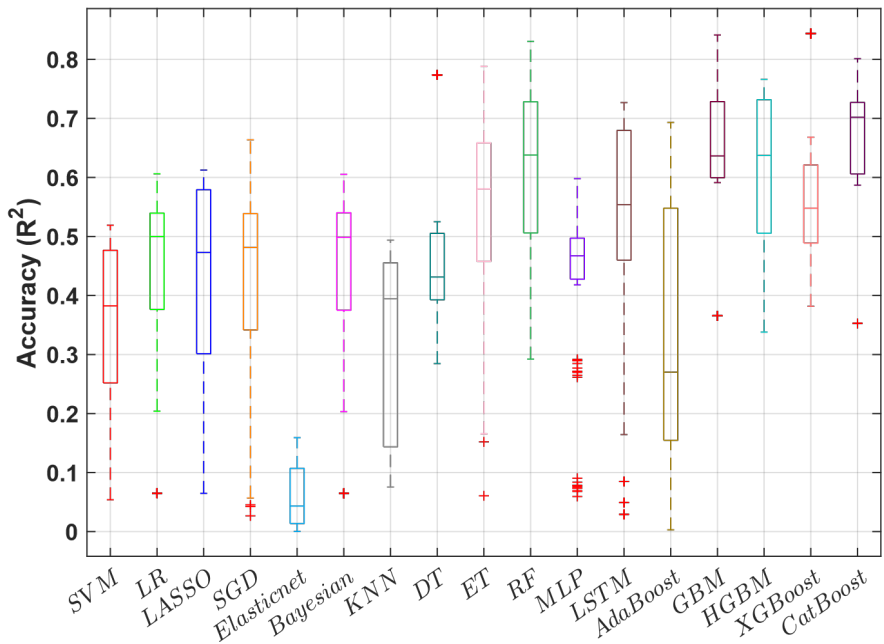

Figure 6: Comparative framework of 17 machine learning models based on their performance ($R^2$) on predicting grinding mill throughput

For a more robust foundation in selecting the top-performing model, we conducted the Friedman Test and subsequent T-test, and the resulting average ranks and p-values are detailed in Table 6. According to these results, CatBoost is the most accurate predictor, achieving the lowest average rank (2.99) and demonstrating a significant difference compared to other models, as indicated by p-values less than the significance level ($\alpha = 0.05$). These results confirm that CatBoost is the superior model for predicting SAG mill throughput. While CatBoost stands out as the best model, it is noteworthy that GBM, HGBM, and XGBoost also exhibited strong and competitive performance. GBM secured the lowest average rank among all remaining models (3.16), showing its effectiveness, while HGBM and XGBoost demonstrated robust performance with average ranks of 4.78 and 6.26, respectively. These findings confirm the effectiveness of ensemble models for predicting SAG mill throughput.

Table 6: Friedman Test and T-test statistical results to indicate the average rank of ML models performance (accuracy=R-value) and the significant difference of the best-performed model (CatBoost) compared with others.

| Mehtod | SVM | LR | LASSO | SGD | ElasticNet | Bayesian | KNN | DT | ET |
|---|---|---|---|---|---|---|---|---|---|
| Average rank (Friedman Test) | 13.05 | 9.63 | 10.99 | 10.71 | 16.89 | 10.17 | 13.27 | 9.93 | 6.86 |
| P-value | 4.10E-69 | 6.63E-35 | 9.72E-33 | 5.86E-35 | 2.94E-73 | 7.04E-35 | 8.08E-50 | 3.80E-20 | 2.27E-13 |
| Method | RF | MLP | LSTM | AdaBoost | GBM | HGBM | XGBoost | CatBoost | |
| Average rank (Friedman Test) | 4.86 | 10.34 | 7.1 | 12.01 | 3.16 | 4.78 | 6.26 | 2.99 | |
| P-value (T-test, $\alpha = 0.05$) | 7.57E-08 | 6.17E-43 | 5.14E-09 | 4.01E-31 | 1.73E-02 | 5.80E-11 | 1.52E-10 | 0.00E+00 | |

*4.2. Outlier detection and feature selection results*

Comparative analysis has identified CatBoost as the most accurate predictive model. In this section, we apply outlier detection and feature selection techniques to assess their impact on enhancing the prediction accuracy of the CatBoost model. Table 7 presents accuracy metrics after removing outliers using the Local Outlier Factor (LOF) technique. While the removal of outliers resulted in a notable 14% improvement in the minimum R2 value, the enhancements for mean and median R2 values are less than 1%. Based on the obtained results, outlier removal will not be applied, as its effect is negligible in this case.

Table 7: Statistical details of CatBoost model after outlier removal using LOF

|  | RMSE | MAE | EVS | MSLE | SMAPE | MAPE | R2 |
|---|---|---|---|---|---|---|---|
| Min | 2.54E+01 | 1.97E+01 | 4.89E-01 | 3.80E-04 | 1.50E+00 | 1.51E+00 | 4.93E-01 |
| Max | 7.80E+01 | 6.36E+01 | 7.76E-01 | 2.48E-02 | 5.35E+00 | 8.33E+01 | 7.87E-01 |
| Mean | 5.16E+01 | 3.79E+01 | 6.45E-01 | 7.76E-03 | 3.13E+00 | 1.55E+01 | 6.71E-01 |
| Median | 4.91E+01 | 3.75E+01 | 6.61E-01 | 2.20E-03 | 3.17E+00 | 3.30E+00 | 7.07E-01 |
| STD | 1.54E+01 | 1.23E+01 | 1.09E-01 | 9.56E-03 | 1.06E+00 | 2.54E+01 | 1.05E-01 |

For feature selection, we employ the recursive feature elimination (RFE) method, as previously mentioned. To determine the optimal number of final features (K), we consider all possible alternatives (from 20 to 1 features) and calculate accuracy metrics. Figure 7 illustrates the mean and median R² values per different K values. As evident, selecting 15 features yields the highest mean and median R² values. Consequently, K=15 features are selected for final modelling. The removed features include %PL 13.2-19mm, %PL 53-75mm, %PL 150-212mm, %PL 212-300mm, and Pebble crusher two working status. Considering the histogram plot of input features as illustrated in Figure 2 shows that removed particle size-related features are mainly sparse features with low variation.

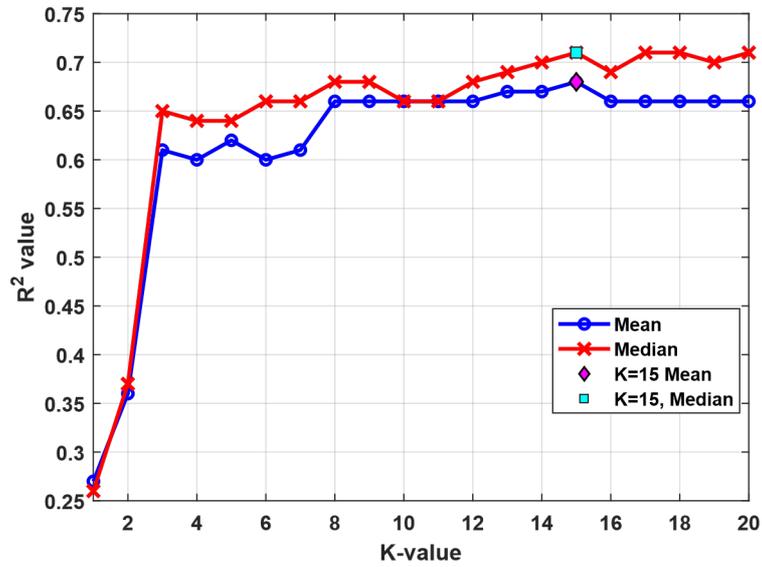

Figure 7: Mean and Median R2 for different number of selected features

After eliminating the aforementioned five less influential features, the CatBoost model is retrained. The best performing model out of 5 runs is selected as the best model to be integrated with optimisation algorithms. This model achieved an R-squared value of 82%. Figure 8 illustrates this model's predictions and the actual mill throughput values.

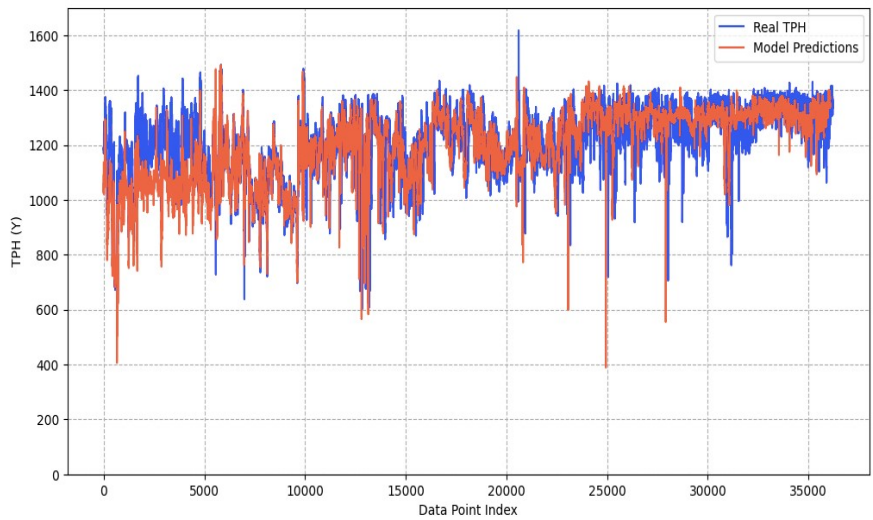

Figure 8: A comparison performance of CatBoost predicted mill throughput with true values.

### *4.3. Optimisation results*

The best CatBoost model is used as the fitness function inside three evolutionary algorithms, including DE, GA, and PSO. Population size and maximum generations are selected as 25 and 50, respectively. All other parameters have been retained at their default values without further optimisation for the specific problem, as presented in Table 8.

Table 8: Optimization algorithms parameters

| Optimisation algorithm | Parameters |
|---|---|
| DE | Crossover Rate=0.7, Differential Weight (F)= 0.5, DE/rand/1 |
| GA | Crossover Rate=0.7, Mutation Rate= 0.2, Mutation Method: Gaussian, Cross Over Method: Two-Point Crossover, Selection Method: Tournament Selection |
| PSO | Velocity Limits ($s_{min}$, $s_{max}$): -1, 1, Cognitive and Social Coefficients ($\phi_1$, $\phi_2$): 2 |

Fifteen real-valued variables are the subject of optimisation, targeting the attainment of maximum SAG mill throughput. Upper and lower bounds for these input features are considered as constraints for optimisation algorithms. Due to the randomness inherent in these optimization algorithms, run results may vary slightly from one execution to another. Therefore, each algorithm was run ten times, and in each run, the highest achieved mill throughput in each generation was recorded. The convergence rate variances of these algorithms are presented in Figure 9 ( to visualize the statistical distribution of the method's performance). In this figure, the colour shade represents the range between the minimum and maximum best mill throughput predictions across all ten runs, while the bold line depicts the average value of these predictions. It is evident that DE exhibits notably lower distribution across multiple runs. Notably, the results demonstrate increasing stability and a reduction in variance as the algorithm progresses from initial evaluations. This trend underscores the robustness and convergence reliability of DE over successive iterations. Remarkably, GA exhibits the highest variability in results among the considered optimisation methods. This finding confirms the comparative stability of DE, emphasizing its potential advantages over other methods in achieving consistent convergence.

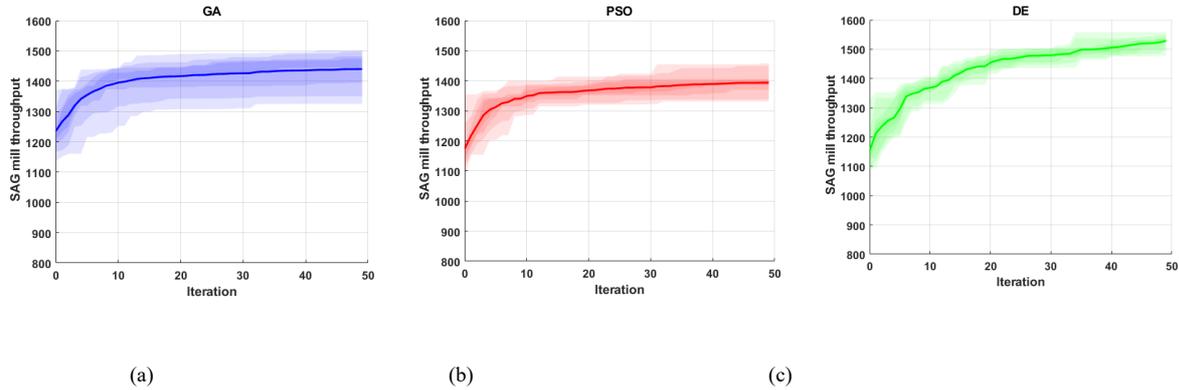

(a)  (b)  (c)

Figure 9: The convergence rate variance of the three optimisation methods to maximise the SAG mill throughput

Figure 10 illustrates a comparative plot of the convergence rates among the employed optimisation algorithms. Each line within the plot represents the average best-so-far solution obtained from ten independent runs. In the comparative analysis of these optimisation algorithms, it can be seen that the DE could show a considerable ability to converge proper solutions and consistently outpaces all other methods, showcasing its superior capability in achieving the highest predicted mill throughput among the evaluated optimisation techniques. Moreover, after a specific generation, approximately 400 and 600 for GA and PSO, respectively, the fitness values exhibit limited improvement. This observation suggests that GA demonstrates the highest convergence speed, followed by PSO. However, when considering the maximum predicted mill throughput as the ultimate metric of success, DE emerges as the leading performer. DE despite PSO exhibiting comparable performance in initial evaluations and GA outperforming DE in this early phase, a noteworthy transition occurs after approximately 400 evaluations, positioning DE as the best-performing method.

It is notable that, although the random uniform method is employed for creating the initial population across all optimization methods, slight differences in the starting points are observable among these approaches. This difference can be attributed to the inherent stochastic nature of the random uniform method. For instance, in the context of Genetic Algorithms (GA), the stochastic generation of individuals across the 10 repeated runs could randomly lead to marginally improved SAG mill throughput predictions.

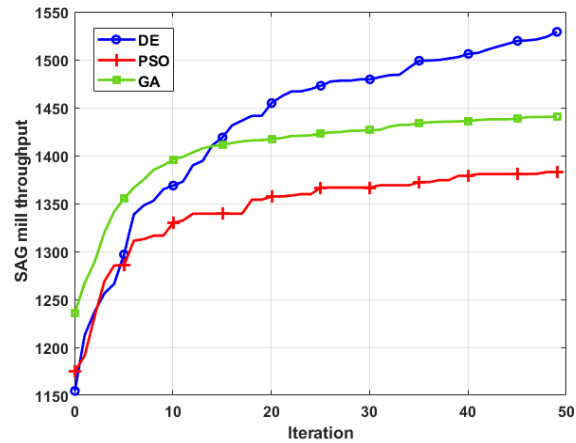

Figure 10: The convergence rate comparison for three optimisation methods: DE, GA, and PSO. Each method runs ten times independently

To have a comprehensive understanding of the solutions in the final generation, Figure 11 presents box plots depicting the best solutions across ten independent runs in this generation. To establish a baseline for assessing the performance of employed optimisation methods, we also sampled 1250 instances using two sampling methods including random uniform and Latin hypercube sampling. Each sampling method was repeated 10 times, and the best solution per each run is determined and depicted as box plots in Figure 11.

Notably, DE consistently emerges as the superior method, exhibiting both higher mill throughput predictions and concurrently demonstrating low variation in results. This outstanding and robust performance strongly positions DE as the superior optimisation algorithm. While GA secures the second position in terms of prediction values, it is noteworthy that its results display a higher degree of variation. PSO yields lower predictions with comparatively lower variance.

The results obtained from all three optimization methods were superior to those obtained from the two employed sampling methods, highlighting the efficacy of optimization techniques in exploring the search space more efficiently compared to simple sampling methods. The comparison between Latin hypercube sampling (LHC) and random uniform sampling (URS) reveals higher variability for LHC, attributed to its systematic and thorough exploration of the entire parameter space, ensuring sampling across a broad range of parameter values. As LHC has effectively explored the search space, it could reach to higher mill throughput predictions.

Based on these results, DE prominently secures the top rank in both solution quality, emphasising high predictions, and result stability. This underscores DE's exceptional performance in achieving superior predictive accuracy while maintaining consistency, solidifying its position as the best optimisation algorithm for maximising SAG mill throughput.

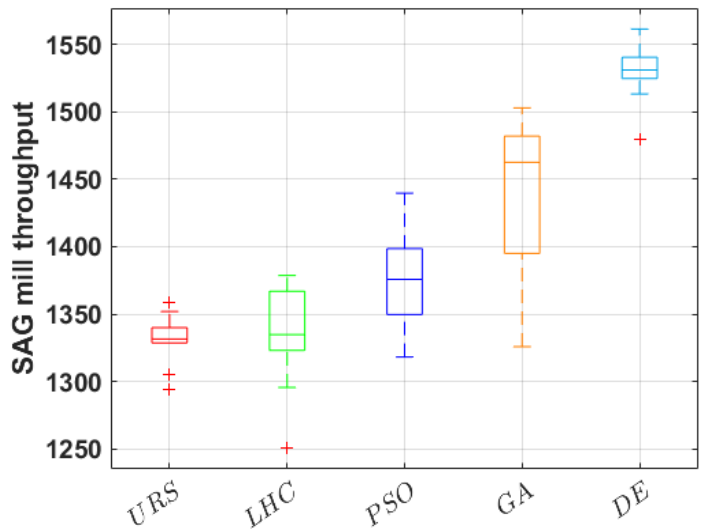

Figure 11: Statistical optimisation results for sampling methods (URS and LHC) and optimisation methods (DE, GA, and PSO)Each method runs ten times independently

Finally, 25 obtained solutions in the last generation of the DE algorithm will be shared with industry experts as potential candidates for maximising SAG mill throughput. These experts will play a pivotal role in selecting the most suitable solution based on practical working conditions. This collaborative approach ensures that the chosen solution is well-aligned with the operational considerations.

## 5. Conclusions

Optimum parameter setting to achieve maximum SAG mill throughput in the mining industry is a critical need. This paper introduces an intelligent framework using expert knowledge, machine learning techniques, and evolutionary algorithms to address this need. For this purpose, an extended industrial data set including 36,743 records is utilised, and effective features are selected according to expert knowledge. After removing erroneous data, 17 machine learning models are compared to predict SAG mill throughput. The most accurate model is selected, and LOF and RFE are employed for outlier detection and feature selection, respectively, to enhance the predictive performance of the selected prediction model. The best model trained with selected features is then integrated with three different optimisation algorithms for finding optimum values of control parameters to achieve maximum SAG mill throughput, considering the working limits of input features as constraints. The optimal solutions, identified through the most effective optimisation algorithm, will be shared with industry experts for informed decision-making based on the current real process conditions.

Our findings identified CatBoost as the most accurate prediction model for estimating SAG mill throughput. Ensemble models, in general, demonstrated strong performance for this specific purpose. In comparing three optimisation models, Differential Evolution (DE) stood out, achieving higher mill throughput predictions and concurrently producing robust and reliable results. Notably, in this application, the implementation of the Local Outlier Factor (LOF) method for outlier detection did not yield substantial improvements, leading us to exclude it from the final modelling. Additionally, Recursive Feature Elimination (RFE) effectively resulted in the removal of five input features. Notably, four of these features were particle size-related variables, predominantly consisting of sparse data points with minimal variations, demonstrating the coherence of the feature selection outcome.

Remarkably, our intelligent framework extends beyond the mere application of ML and AI methods. The engagement of an expert in selecting relevant features and evaluating proposed solutions ensures an informed optimisation process. One of the distinguishing characteristics of our framework lies in its adaptability to the dynamic nature of the grinding process. In contrast to static models, our model is capable of real-time updates, enabling finely tuned modelling that perfectly aligns with the changing conditions of the grinding process. This dynamic capability not only ensures the model's accuracy but also supports its secure and efficient long-term use.

For further developments, incorporating more inputs, in particular indicators of ore hardness, presents the potential for augmenting predictive accuracy. It is also promising to conduct hyperparameter tuning to enhance the performance of prediction and optimisation methods. Additionally, the hybridisation of different algorithms is a practical approach to leverage the strengths inherent in each method.

**Data availability**

The data that has been used is confidential.

**Acknowledgements**

This research was supported by the Australian Research Council Integrated Operations for Complex Resources Industrial Transformation Training Centre (project number IC190100017) and funded by universities, industry and the Australian Government.